\documentclass[a4paper,12pt]{report}

\usepackage[pdfusetitle,bookmarksnumbered,plainpages=false]{hyperref}

\usepackage[british]{babel}
\usepackage[a4paper,top=2.5cm,bottom=2.5cm,left=4cm,right=2cm,headsep=10pt]{geometry}
\usepackage{fancyhdr}
\fancypagestyle{plain}{
	\fancyhf{}
    \lhead{\textit{30th August 2012}}
    \chead{\thepage}
	
}
\usepackage{latexsym}
\usepackage{amsmath}
\usepackage{textcomp}
\usepackage{amssymb}
\usepackage{rotating}

\begin{document}

\begin{titlepage}

\begin{center}
\huge\textbf{An Analysis On Ward Identity For Multi-Field Inflation}
\vskip5mm
\Large\textbf{Candidate Number: 33161}
\normalsize
\end{center}
\vfill
\begin{flushleft}
\large
Submitted for the degree of Master of Science \\
University of Sussex \\
30th August 2012
\end{flushleft}
\end{titlepage}

\pagenumbering{roman}
\chapter*{Summary}

Given a correlation function ( or n-point function ), can the corresponding nature of space-time be determined ? To answer this question it is required to derive the Ward Identity ( WI ), analyse the symmetries and arrive at the law of conservation. \textit{Modus operandi} involves Lie differentiating two-point function considering the symmetry to be non-anomalous. The WI so obtained is shown to form a Lie algebra which determines the nature of space-time. Solving the identity results in a law of conservation, which physically explains the reason for WI to form an algebra and contains in it an equation of motion for four-point function. As a special case, a relation between mass and potential involving the spatial derivatives of four- and five- point function is obtained. Finally, the conservation equation is exploited to get the probability amplitude for the two-point function which shows how correlation functions provide an opportunity to probe the fundamental laws of physics.

\newpage

\chapter*{Preface}

The dissertation work is divided into six chapters with emphasis on the derivation of Ward Identity ( WI ) for multi-field inflation from the first principles, using path integral formulation and then subject it to analysis for arriving at satisfactory results. 
\begin{itemize}
\item Chapter 1 motivates the framework - Inflation.
\item Chapter 2 gives a pedagogical approach to path integral formulation and derivation of WI.
\item Chapter 3 explains de Sitter space and its properties.
\item Chapter 4 involves the derivation and analysis of WI for Lagrangian density pertaining to multi-field inflation.
\item Chapter 5 discuss the results obtained from the analysis. 
\item Chapter 6 gives the conclusion.
\end{itemize}
Chapters 1, 2 and 3 rely on literature for explaining the background material required for calculations shown in chapter 4, which I have carried out along with my supervisor followed by discussion of results and conclusion in the last two chapters respectively.
\newpage
\chapter*{Conventions}
As the work depends on quantum field theory and general relativity certain conventions are fixed from the outset. Correlation function in quantum field theory is defined as
$$ \langle \Omega |T\{\phi(x_{1})\phi(x_{2})\cdots\phi(x_{n})\} | \Omega\rangle $$
where \(T\) denotes the time-ordering and \(|\Omega\rangle\) the ground state. As the time-ordering will be maintained explicitly till the end, correlation functions will appear without the time-ordering symbol. For example, the two-point function will be written as \(\langle \phi(x_{1})\phi(x_{2})\rangle\). The coordinates chosen are \((\eta, x, y, z)\), where \(\eta\) denotes the conformal time. The calculations are done on Robertson-Walker metric for spatially flat condition \((K = 0)\), with unit \( c = 1 \). The metric is given by
$$ ds^{2} = a^{2}(\eta) (d\eta^{2} - dx^{2} - dy^{2} - dz^{2})$$
and the metric tensor is expressed as 
$$ g_{\mu\nu} = diag(a^{2}(\eta), -a^{2}(\eta), -a^{2}(\eta), -a^{2}(\eta)) $$
The fields in our case are quantum fluctuations, denoted by \(\delta\phi(\eta, \bf{x})\), where \(x^{\mu} \equiv (\eta, \bf{x})\) is a four-vector and \(\bf{x}\) is the three dimensional spatial vector. \\
\\
Differentiation with respect to conformal time will denoted by a prime, like \(a'\) and by a dot, like \(\dot{a}\) with respect to time.

\newpage
\pdfbookmark[0]{Contents}{contents_bookmark}
\tableofcontents
\listoftables
\phantomsection
\addcontentsline{toc}{chapter}{List of Tables}

\newpage
\pagenumbering{arabic}

\chapter{The Paradigm}
\label{chap:inflation}

Universe is populated with massive structures bound gravitationally. The building blocks of these massive structures are the elementary particles belonging to the Standard Model and yet to be determined entities beyond the Standard Model. Apart from the observed bodies on large scales, cosmos has many intriguing aspects in terms of matter, energy and interactions. Observations and analysis done on these leads to the fundamental question - what is the source for this evolution ? One suitable source is the quantum fluctuations that would have existed in the very early Universe. The fluctuations lead to density perturbations which is responsible for the formation of massive structures. These quantum fluctuations are treated as fields since they belong to the early Universe, where the energies involved are very high. It is required to understand their dynamics and hence determine the background field and before we could do that, a framework has to be chosen which supports the existence of quantum fluctuations. The paradigm is called as inflation.
\paragraph{}
Fundamentals of inflationary cosmology\cite{lecture} were first developed in the early 1970's by considering energy density of a scalar field to be a cosmological constant, changing during the phase transitions. In 1978 Andrei Linde and Gennady Chibisov proposed their idea which was based on exponential expansion of the Universe, but was dropped due to certain shortcomings. In 1979 Alexei Starobinsky developed a model based on conformal anomaly, by assuming the Universe to be homogeneous and isotropic from the beginning. Despite of its failure in being an inflationary model, it succeeded in predicting the gravitational waves with a flat spectrum. In fact, the mechanism proposed by Mukhanov and Chibisov which explained the production of adiabatic perturbations of the metric with a flat spectrum was based on Starobinsky model. Finally in 1981 Alan Guth developed the framework, which is currently called as \textquotedblleft old inflation \textquotedblright, based on supercooling during the cosmological phase transitions. It is this model which played a substantial role in the development of inflationary cosmology. Guth work was followed by \textquotedblleft new inflation \textquotedblright developed by Albrecht and Steinhardt, and independently by Linde in 1982, \textquotedblleft chaotic inflation \textquotedblright in 1983 by Linde and in these thirty years, this paradigm has evolved to such an extent that inflationary models based on superstring theory are being developed. Inflationary cosmology has surpassed the levels of being a candidate that satisfies the shortcomings of hot big bang model and has grown into a theory that is capable of explaining the origin of structure in the Universe.

\section{Shortcomings}

The standard hot big bang model\cite{lid} successfully explains the  expansion of the Universe, existence of the cosmic microwave background, nucleosynthesis of the light elements, growth of structures in the Universe due to gravitational collapse and predicts the age of the Universe. These are crucial observational tests qualified by the hot big bang model, but still was not capable of describing the Universe at earlier epochs when the conditions very hot and energies involved are very high. Its success was limited to the epochs when the conditions prevailing in the Universe was cool and the dynamics were based on physics that we could observe from various experimental methods known to us. The success of the hot big bang model was so obvious that, its shortcomings needed to be rectified without affecting it. 
\paragraph{}
 
The standard hot big bang model\cite{guth,shinji} had to address the initial conditions, in order to explain very early time periods of the Universe, which will subsequently lead to the present conditions described by the big bang model. The initial conditions are :
\begin{enumerate}
\item Horizon problem: The early Universe is assumed to be homogeneous, but still contains causally disconnected regions. Hence, dynamics would have violated causality. To grasp this, consider the physical wavelength \(a\lambda\) well inside the Hubble radius \(H^{-1}\). In hot big-bang scenario the physical wavelength and Hubble radius evolves as \(a\lambda \propto t^{p}\) where \((0 < p < 1)\) and \(H^{-1} \propto t\) respectively. Hence, physical wavelength tends to be smaller than Hubble radius as time passes and regions were causality is valid becomes a small fraction of Hubble radius. Microwave background photons are emitted at the time of decoupling. Let \(t_{dec}\) be the time of decoupling, \(D_{H}(t_{dec})\) and \(d_{H}(t_{dec})\) be the particle horizon and comoving distance respectively, such that \(D_{H}(t_{dec})=a(t_{dec})d_{H}(t_{dec})\) where the particle horizon represents the region in which the dynamics involved were causally valid. From observations it has been calculated that,
\begin{equation}
\frac{d_{H}(t_{dec})}{d_{H}(t_{0})} \approx 10^{-2}
\end{equation}
where \(t_{0}\) is the present time. This indicates that the regions where photons could contact causally at that time is very small. It has been calculated that the surface of last scattering corresponds to an angle of order \(1^{o}\). But observations reveal photons in thermal equilibrium and at same temperature in all regions in the Cosmic Microwave Background sky. 
\item Flatness problem: According to Friedmann equation,
\begin{equation}
\label{eq:man}
| \Omega_{tot} - 1 | = \frac{|K|}{a^{2}H^{2}}
\end{equation}
where 
\begin{equation}\label{eq:ev}
a^{2}H^{2} \ \propto t^{-1},\ a^{2}H^{2}\ \propto t^{-2/3}
\end{equation}
hence,
\begin{equation}\label{eq:tn}
|\Omega_{tot} - 1 | \propto t,\ |\Omega_{tot} - 1| \propto t^{2/3}
\end{equation}
for radiation and matter domination respectively. In the standard big-bang model \(a^{2}H^{2}\) decreases, due to which value of \(\Omega_{tot}\) drifts from unity as Universe expands, whereas observations show that its value is in the order of magnitude of one. From \eqref{eq:ev}-\eqref{eq:tn} it is seen that the difference between \(\Omega_{tot}\) and 1 increases with respect to time, which means that flat geometry is an unstable condition for the Universe and any deviation will lead to its collapse. Hence, Universe to be flat as observed now, its geometry should have been close enough to being flat during the very early time periods, for which the initial value of the Hubble constant should have been extremely fine-tuned.
\item Monopole problem: Monopoles are unwanted relics produced by the breaking of supersymmetry. Their energy density decrease in the order of \(a^{-3}\). As energy density of radiation decreases in the order of \(a^{-4}\) during the era of radiation-domination, these relics should have populated the Universe but contradicts current observations.
\item Observations done by COBE satellite reveal anisotropies in the surface of last scattering with amplitudes being small and close to scale-invariance. But in the standard scenario it is not possible for these fluctuations to be generated between big bang and last scattering as they have to spread out on a very large scale. Hence the hot big bang model cannot explain the origin of large-scale structure.
\end{enumerate}
These shortcomings are resolved by the inflationary paradigm.

\section{Inflation}
The basic idea of inflation\cite{lidintro} is 
\begin{equation}
\ddot{a} > 0
\end{equation} 
an epoch when the expansion of the Universe is accelerating. The underlying idea is to separate the causal size from the Hubble radius, so that the real size of
the horizon region in the standard radiation dominated era is larger than the
Hubble radius, which implies that the comoving Hubble radius decreases
rapidly in the very early universe. 
\begin{equation}
\frac{d(H^{-1}/a)}{dt} < 0
\end{equation}
It should be noted that inflation is an epoch during the early time period of the evolution and should come to an end paving way for the standard hot big bang theory and its success. 
\subsection{Resolving the Shortcomings}
\begin{enumerate}
\item Horizon problem: During inflation \(p > 1\), corresponding to which the physical wavelength \(a\lambda\) grows faster than the Hubble radius \(H^{-1}\) as a result of which physical wavelength remains outside the Hubble radius. Hence, the region where causality is valid gets stretched on scales much larger than the Hubble radius. This solves the horizon problem.
\item Flatness problem: By virtue of definition, \(a^{2}H^{2}\) in \eqref{eq:man} increases during inflation due to which \(\Omega\) tends to unity and the flatness problem is solved.
\item Monopole problem: The unwanted relics get diluted by the accelerated expansion, as the energy density of the Universe decreases slowly (in the order of \(a^{-\frac{2}{p}}\)) than the energy density of the relics. However these massive relics might be produced after inflation. The process of reheating ensures that the energy density of the Universe gets converted into matter and radiation, provided the temperature is low enough to avoid the formation of unwanted relics.
\item During inflation the comoving Hubble radius decreases which leads to the generation of density perturbations scale-invariant on large scales. During the inflationary phase the scales of perturbation are inside the Hubble radius, which allows causality to generate the quantum fluctuations. After the first horizon crossing perturbations become classical and by end of inflation comoving Hubble radius increases, followed by the second horizon crossing which makes the causality valid. Therefore, small perturbations generated during inflation are observed as large-scale perturbations after the second horizon crossing. Inflation successfully provides a mechanism valid causally in generating the density perturbations observed in the microwave background anisotropies.
\end{enumerate}
Following its success in explaining initial value problems, inflation has now become a theory capable of explaining large-scale structure of the Universe.

\section{Dynamics}
Given the framework, it is required to have an appropriate structure of dynamics which determines its evolution. The motivation\cite{brand} for inflation was based on predictions from particle physics, naturally satisfying the requirements for the very early conditions of the Universe, where energy was in the order of 1 GeV or more. Under such conditions matter is in a highly compressed form and theory of quantum fields is capable of giving a robust structure to the inflationary dynamics, with scalar fields \((\phi)\) as the primary object under evolution. These scalar fields depend on space-time and play the role of operators. Given a scalar field, Lagrangian density is constructed which is made out of kinetic and potential terms.  The action is defined as,
\begin{equation}
\mathcal{S} = \int d^{4}x \mathcal{L}(\partial_{\mu}\phi, \phi)
\end{equation}
which maps functions to numbers - \textit{functional}, hence the action principle states that
\begin{equation}
\label{eq:action}
\delta\mathcal{S} = 0
\end{equation}
which leads to the Euler-Lagrange equations of motion,
\begin{equation}
\label{eq:euler}
\partial_{\mu}\left(\frac{\partial\mathcal{L}}{\partial\left(\partial_{\mu}\phi\right)}\right) - \frac{\partial\mathcal{L}}{\partial\phi} = 0
\end{equation}
As a generic example\cite{pert} consider Einstein-Hilbert action,
\begin{equation}
\mathcal{S} = \int d^{4}x \sqrt{-g}\left(\frac{1}{2}g^{\mu\nu}\partial_{\mu}\phi\partial_{\nu}\phi - V(\phi)\right)
\end{equation}
where \(g_{\mu\nu} = diag\left(1, -a^{2}(t), -a^{2}(t), -a^{2}(t)\right)\). From \eqref{eq:action} and \eqref{eq:euler}, we get
\begin{equation}
\label{eq:ex}
\ddot{\phi} + 3H\dot{\phi} - \frac{1}{a^{2}}\nabla^{2}\phi + \frac{d V(\phi)}{d\phi} = 0
\end{equation}
If the field is homogeneous, then \(\nabla^{2}\phi = 0\). For n-number of fields, \eqref{eq:ex} can be generalised as 
\begin{equation}
\ddot{\phi_{n}} + 3H\dot{\phi_{n}} - \frac{1}{a^{2}}\nabla^{2}\phi_{n} + \frac{\partial V}{\partial \phi_{n}} = 0
\end{equation}
In case of a homogeneous scalar field in an expanding universe, where the energy and pressure densities are given by 
\begin{equation}
\rho = \frac{1}{2}\dot{\phi}^{2} + V(\phi)
\end{equation}
\begin{equation}
p = \frac{1}{2}\dot{\phi}^{2} - V(\phi)
\end{equation}
from the Friedmann equation and \eqref{eq:ex} we get
\begin{equation}
H^{2} = \frac{8\pi}{3}\left[ \frac{1}{2}\dot{\phi}^{2} + V(\phi)\right]
\end{equation}
\begin{equation}
\ddot{\phi} + 3H\dot{\phi} = -\frac{dV}{d\phi}
\end{equation}
where Planck mass \(m_{Pl}\), is set to unit value. During inflation \(\dot{\phi}^{2} < V(\phi)\), hence potential energy dominates. Various inflationary models are proposed depending on the choices of \(V(\phi)\), but still its exact nature during the inflationary epoch remains unclear to the scientific community. It is stated that the potential should be almost flat and have a minimum in which inflation can end.
\section{Model}
Observations\cite{mf} on large-scale structure, CMB, supernovae and other such astronomical bodies, support the inflationary framework. But it is required to have a realistic model and of various candidates, multi-field inflation is a preferable choice. As the name suggests, more that one scalar field and sometimes many of them involved are in the dynamics offering higher degrees of freedom. The motive behind the choice of multi-field inflation is the situation that would have existed during the inflationary epoch involving very high energies, very dense and hot conditions, which naturally demands the presence of more than one fields in the scenario. Apart from this in order to incorporate theories beyond Standard Model, presence of multiple-fields is essential. This is will naturally make inflation a robust theory.

\chapter{Formalism}
A pedagogical\cite{tak,pes} overview of path integral formalism and Ward Identity ( WI ) is given in Section 2.1 and 2.2 respectively. 

\section{Path Integral Formulation}
Path integral formalism ( or functional integral ) was proposed by Richard P. Feynman for calculating the amplitude of propagation, for a quantum mechanical system from one point in space to another point in a given time. In the Hamiltonian formulation,
\begin{equation}
\label{eq:path}
U(x_{a}, x_{b}; T) = \langle x_{b} | e^{-iHT/ \hbar} | x_{a} \rangle
\end{equation}
where \(U\) is the amplitude, \(x_{a}, x_{b}\) and \(T\) are the points is space and given time respectively. The amplitude corresponds to a non-relativistic quantum-mechanical system, in the position representation of the Schr\"{o}dinger time-evolution operator. An important advantage of quantum mechanics, which lacks in classical physics is the principle of superposition. According to this principle, if a system can evolve between the initial and final points in more than one way, then the total amplitude for the process is just the sum of individual amplitudes. Taking advantage of superposition principle, in the path integral the expression for \(U\) is not like \eqref{eq:path}, instead
\begin{equation}
\label{eq:u}
U(x_{a}, x_{b}; T) = \sum_{all\ paths} e^{i\ (phase)} = \int \mathcal{D}x(t)\ e^{i\ (phase)}.
\end{equation}
where \(\int\mathcal{D}x(t)\) indicates \textit{sum over all paths}- an integral over continuous space of functions. The integrand in \eqref{eq:u} associates a complex amplitude corresponding to the function \(x(t)\), thereby mapping functions to numbers. Hence, it is a functional and the symbol \(\mathcal{D}\) represents the measure of functional or path integral. An appropriate choice for the phase is action \(S\), since it determines the path taken by the system through the method of stationary phase. Therefore amplitude of propagation is given by
\begin{equation}
U(x_{a}, x_{b}; T) = \int \mathcal{D}x(t)\ e^{iS[x(t)]/\hbar}
\end{equation}
where \(\hbar\) is to ensure the validity of method of stationary phase in the classical limit. Let us set \(\hbar = 1\) and generalise path integrals as obtained for a non-relativistic quantum mechanical system to scalar fields in quantum field theory. The procedure is straight forward, 
\begin{equation}
x(t) \rightarrow \phi(\vec{x}, t)
\end{equation}
and the amplitude is 
\begin{equation}
U(\phi_{a}, \phi_{b}; T) = \int \mathcal{D}\phi\ e^{i\mathcal{S}[\phi]}
\end{equation}
where \(\mathcal{S} = \int d^{4}x\ \mathcal{L}\), with the boundary conditions
\begin{eqnarray}
\phi = \phi_{a}(\vec{x}) :\  t = T_{a}\\
\phi = \phi_{b}(\vec{x}) :\  t = T_{b} 	\nonumber
\end{eqnarray}
\paragraph{}
With the the help of path integrals and the above mentioned boundary conditions, we construct a mathematical object
\begin{equation}
\label{eq:corr}
C = \int \mathcal{D}\phi\ \phi(x_{1})\phi(x_{2})e^{i\mathcal{S}[\phi]}
\end{equation}
such that at times \(t_{1}\) and \(t_{2}\), the field configurations are
\begin{equation}
\phi(\vec{x},t_{1}) = \phi_{1}(\vec{x}), \ \phi(\vec{x},t_{2}) = \phi_{2}(\vec{x}) 
\end{equation}
and integrating over \(\phi_{1}\) and \(\phi_{2}\) we get a functional integral segmented as
\begin{equation}
\int \mathcal{D}\phi = \int \mathcal{D}\phi_{1}\int\mathcal{D}\phi_{2}\int\mathcal{D}\phi\  |_{\phi_{1}(\vec{x}),\phi_{2}(\vec{x})}
\end{equation}
where
\begin{equation}
\label{eq:main}
\int\mathcal{D}\phi\ |_{\phi_{1}(\vec{x}),\phi_{2}(\vec{x})}\ e^{i\mathcal{S}[\phi]} = 
\langle\phi_{b}|e^{-iH(T_{b}-t_{2})}|\phi_{2}\rangle \langle\phi_{2}|e^{-iH(t_{2}-t_{1})}|\phi_{1}\rangle \langle\phi_{1}|e^{-iH(t_{1}-T_{a})}|\phi_{a}\rangle
\end{equation}
with \(t_{1} < t_{2}\) determining the order. In the Schr\"{o}dinger and Heisenberg picture
\begin{equation}
\label{eq:pic}
\hat{\phi}_{S}(\vec{x}_{1,2})|\phi_{1,2}\rangle = \phi_{1,2}(\vec{x}_{1,2})|\phi_{1,2}\rangle , \  \hat{\phi}_{H}(\vec{x},t) = e^{iHt}\hat{\phi}_{S}(\vec{x})e^{-iHt}\ .
\end{equation}
Making use of \eqref{eq:main}-\eqref{eq:pic} in \eqref{eq:corr} and repeating the exercise for \(t_{2} < t_{1}\) followed by addition of the expressions obtained, we get
\begin{equation}
C = \langle\phi_{b}|e^{-iHT}\ T\{\hat{\phi}_{H}(x_{2})\hat{\phi}_{H}(x_{1})\}\ e^{-iHT} |\phi_{a}\rangle\ .
\end{equation}
By making an appropriate choice for the limit of $T$, such that
\begin{equation}
e^{-iHT}|\phi_{a}\rangle = \lim_{T \rightarrow \infty(1 - \epsilon)} \sum_{n} e^{-iHT}|n\rangle\langle n|\phi_{a}\rangle = |\Omega\rangle\langle\Omega|\phi_{a}\rangle e^{-iE_{0}T}
\end{equation}
which projects out the vacuum state $|\Omega\rangle$ from $|\phi_{a}\rangle$ and $|\phi_{b}\rangle$, one can get rid of all the $n \neq 0$ terms, i.e., all states with energy above vacuum, as it decays exponentially fast in limit \(T \rightarrow \infty(1 - \epsilon)\) and the only survival term in the series is for $E_{0} \equiv \langle\Omega|H|\Omega\rangle$. By normalisation, the factors and phases cancel out and we get the result as

\begin{equation}
\label{eq:def}
\langle\Omega|T\{\hat{\phi}_{H}(x_{2})\hat{\phi}_{H}(x_{1})\}|\Omega\rangle = \lim_{T \rightarrow \infty(1 - \epsilon)} \frac{\int\mathcal{D}\phi\ \phi(x_{1})\phi(x_{2})\ e^{i\mathcal{S}}}{\int\mathcal{D}\phi\ e^{i\mathcal{S}}}
\end{equation}
This result is generalised for n-number of fields 
\begin{equation}
\langle\Omega|T\{\hat{\phi}_{H}(x_{1}) \cdots \hat{\phi}_{H}(x_{n})\}|\Omega\rangle = \lim_{T \rightarrow \infty(1 - \epsilon)} \frac{\int\mathcal{D}\phi\ \phi(x_{1})\cdots \phi(x_{n})\ e^{i\mathcal{S}}}{\int\mathcal{D}\phi\ e^{i\mathcal{S}}}
\end{equation}
and is defined as correlation or n-point functions. By virtue of their definition, they reflect causality. One can perform calculations using the path integral representation of correlation functions and still arrive at the same results as obtained in canonical quantisation, given the fact that the formalism is manifestly Lorentz-invariant. Moreover, there is no need to invoke the perturbation theory, which makes the analysis complicated. 
\section{Ward Identity}
With the correlation functions at our disposal we can exploit it to understand the rudiments of a theory, like in our case inflation relied on the predictions of particle theory and it is well known that symmetry plays a crucial role in it. In fact symmetry is a key factor in the dynamics of a theory, as N\"{o}ether theorem states that every symmetry has a law of conservation associated with it. But N\"{o}ether theorem is restricted to classical field theory and in the realm of quantum fields Ward Identity ( WI ) takes up the role in analysing the symmetries. It should be noted that WI is not a mere extension of N\"{o}ether theorem in the theory of quantum fields, instead is a \textit{sine qua non} in performing model-independent analysis. Its potent was first realised in relating the three point vertex function to the fermion two point function at the tree level in Quantum Electrodynamics ( QED ). This celebrated result,
\begin{equation}
-i(p-q)^{\nu}S_{0}(p)\varGamma_{\nu}(p;q)S_{0}(q) = S_{0}(p) - S_{0}(q)
\end{equation}
is the Ward-Takahashi identity, a consequence of gauge invariance of QED Lagrangian density.
\paragraph{}
We arrive at a generic representation of WI as follows. Consider a Lagrangian, 
\begin{equation}
\label{eq:lag}
\mathcal{L} = \partial_{\mu}\phi\partial^{\mu}\phi^{*} - m^{2}\phi\phi^{*}
\end{equation}
The Lagrangian has a symmetry
\begin{equation}
\phi \rightarrow \phi^{'} = e^{i\alpha}\phi
\end{equation}
Using the Taylor series expansion for the exponential and neglecting the higher order terms,
\begin{equation}
\phi^{'} = (1 + i\alpha)\phi
\end{equation}
hence, infinitesimally
\begin{equation}
\delta\phi = i\alpha\phi
\end{equation}
which transforms \eqref{eq:lag} as
\begin{equation}
\delta\mathcal{L} = i\ \partial_{\mu}\alpha ( \phi\partial^{\mu}\phi^{*} - \phi^{*}\partial^{\mu}\phi )
\end{equation}
where \(j^{\mu}(x)=i\ ( \phi\partial^{\mu}\phi^{*} - \phi^{*}\partial^{\mu}\phi )\) is the N\"{o}ether current, such that \(\partial_{\mu}j^{\mu} = 0\) from the equations of motion. Due to the invariance of the measure \(\mathcal{D}\)
\begin{equation}
\int \mathcal{D}\phi\mathcal{D}\phi^{*}\ \phi_{1}\dots \phi_{n}\ e^{i\mathcal{S}[\phi,\phi^{*}]} = \int \mathcal{D}\phi\mathcal{D}\phi^{*}\ \phi^{'}_{1} \dots \phi^{'}_{n}\ e^{i\mathcal{S}[\phi^{'},\phi^{'*}]}
\end{equation}
and expanding to first order in \(\alpha\)
\begin{equation}
0 = \int \mathcal{D}\phi\mathcal{D}\phi^{*}\ e^{i\mathcal{S}[\phi,\phi^{*}]}\left(i\ \int d^{4}y(\partial_{\mu}\alpha(y))j^{\mu}(y)\phi_{1}\dots\phi_{n}+\sum_{i=1}^{n}\phi_{1}\dots\delta\phi_{i}\dots\phi_{n}\right)
\end{equation}
followed by partial integration and dropping out surface terms we get
\begin{equation}
0 = \int \mathcal{D}\phi\mathcal{D}\phi^{*}\ e^{i\mathcal{S}[\phi,\phi^{*}]}\left((\partial_{\mu}j^{\mu}(y))\phi_{1}\dots\phi_{n})+i\sum_{i=1}^{n}\phi_{1}\dots(\pm i)\delta(y-x_{i})\phi_{i}\dots\phi_{n}\right)
\end{equation}
In bracket notation,
\begin{equation}
\langle\partial_{\mu}j^{\mu}(y)\ \phi(x_{1})\dots\phi(x_{n})\rangle = -i\sum_{i=1}^{n}\langle\phi(x_{1})\dots(\pm i)\delta(y-x_{i})\phi_{i}\dots\phi(x_{n}) \rangle\ .
\end{equation}
This is known as Ward Identity ( WI ) in quantum field theory, relying on the conservation of current and gauge invariance of Lagrangian. In this derivation the time-ordering is assumed and Lagrangian being free of interaction terms, the transition amplitude is between vacuum states. 
\chapter{De Sitter Space}
Methods of quantum field theory\cite{burg,lasenby} in de Sitter space occupies an important place in inflationary cosmology as it provides a scenario where Universe was approximately de Sitter for certain time. The scalar fields in de Sitter space give rise to vacuum energy that leads to inflation. Currently there is compelling evidence for the existence of cosmological constant or some form of an unknown component dubbed as dark energy which is thought to be a relic of symmetry breaking in the early universe owing its origin to a theory depending on field theoretic methods on de Sitter space. Apart from the physical motivation, its nature of being maximally symmetric is another reason for its importance in this paradigm. 
\section{Definition}
A \textit{maximally symmetric} space is an \(n\)-dimensional manifold\cite{car} with \(n\ (n+1)/2\) Killing vectors, with same curvature everywhere and in every direction expressed by translational and rotational isometry respectively. Killing vectors are generators of these isometry; vector fields along which the metric is preserved. Such a manifold is characterised by the scalar quantity derived from the curvature tensor called as Ricci scalar \(R\) which remains same throughout the manifold. For a maximally symmetric space
\begin{equation}
R_{\mu\nu\sigma\delta} = \frac{R}{n(n-1)}\ \left(g_{\mu\sigma}g_{\nu\delta}-g_{\mu\delta}g_{\nu\sigma}\right)
\end{equation}
and the normalised Ricci curvature \(\kappa\) is given by
\begin{equation}
\kappa = \frac{R}{n(n-1)}
\end{equation}
\textit{De Sitter} is maximally symmetric with a constant negative curvature\footnote{Depending on sign conventions it can also be defined with positive curvature.} \((R < 0)\). As a manifold it is Lorentzian (metric is not positive definite) and analogous to non-Euclidean geometry. When \(n = 4\), it pertains to general relativity as a vacuum solution to Einstein equation with a positive cosmological constant.
\section{Symmetry Properties}\label{sec:sym}
Presence of symmetries\cite{rog} leads to the usage of group theory. Space-time manifolds are characterised by continuous groups called as \textit{Lie Groups}, corresponding to which there exist \textit{Lie Algebra} which involves the commutation relations of the generators of the group as follows
\begin{equation}
[X_{i}, X_{j}] = i f^{k}_{ij}X_{k}
\end{equation}
where \(X\) is the generator and \(f^{k}_{ij}\) are the structure constants. Since, metric plays a vital role in the dynamics, the symmetry involved will be reflected in its invariance under transformations. A group that preserves the metric under transformation is called as an orthogonal group \(O(p,q)\), where \(p\) and \(q\) determine the signature of the metric. If determinant of the metric is 1 $(|g_{\mu\nu}| = 1) $ then it is called as special orthogonal group \(SO(p,q)\) which is a Lie group, whose generators are Killing vectors and commutation relation among them forms the associated Lie algebra \(so(p,q)\).
\paragraph{}
Cosmological models demand Robertson-Walker space-time\cite{so}  which are conformally flat and conformally\footnote{Transformations of the form $g_{\mu\nu} \rightarrow \Omega^{2}\ g_{\mu\nu}$} invariant under the action of Lie group $SO(4,2)$, which is the conformal group of Minkowski space-time. Robertson-Walker space-time are spatially homogeneous and isotropic and for the case of $n=4$ with constant negative curvature with $(R < 0)$, it is de Sitter space-time with 10 linearly independent Killing vectors, being maximally symmetric. Being the Lorentzian analogue of non-Euclidean geometry, de Sitter space-time has the Lie algebra $so(3,1)$ corresponding to the Lorentz subgroup $SO(3,1)$ naturally in it. The de Sitter group $SO(4,1)$ is a subgroup of the conformal group $SO(4,2)$, hence the associated Lie algebra $so(4,2)$ contains in it the sub algebras 
\begin{equation}\label{eq:so}
so(4,2) \supset so(4,1) \supset so(3,1) \supset so(3)
\end{equation}
\chapter{Derivation and Analysis}
Ward Identity ( WI ) for a two-point function is derived. WI is then subjected to analysis and solved for arriving at satisfactory conclusions. 
\section{Derivation of Ward Identity}
Consider a generic Lagrangian density pertaining to  multi-field inflation for the quantum fluctuations in Robertson-Walker space-time
\begin{equation}\label{eq:inf}
\begin{split}
\mathcal{L} =   \frac{a^{2}}{2} & \Bigl[\partial_{\eta}\delta\phi_{\lambda}\partial^{\eta}\delta\phi_{\lambda} -\partial_{i}\delta\phi_{\lambda}\partial^{i}\delta\phi_{\lambda} - a^{2}\ m_{\kappa\omega}\ \delta\phi_{\kappa}(\eta,\textbf{x})\delta\phi_   {\omega}(\eta,\textbf{x}) - \\
                                & \frac{a^{2}}{3}\ V_{\kappa\omega\zeta}\ \delta\phi_{\kappa}(\eta,\textbf{x})\delta\phi_{\omega}(\eta,\textbf{x})
               \delta\phi_{\zeta}(\eta,\textbf{x})    \Bigr]
\end{split}
\end{equation}
where \(\lambda\) denotes summation over the fields involved and the action is 
\begin{equation}\label{eq:ac}
\mathcal{S} = \int d^{3}x\ d\eta\ \mathcal{L}
\end{equation}
where \(\mathcal{L}\) is given by \eqref{eq:inf}. Consider a two-point function 
\begin{equation}\label{eq:cor}
C((\eta_{1},\textbf{x}_{1}),(\eta_{2},\textbf{x}_{2})) := \langle\delta\phi_{\alpha}(\eta_{1},\textbf{x}_{1})\delta\phi_{\beta}(\eta_{2},\textbf{x}_{2})\rangle
\end{equation}
where 
\begin{equation}\label{eq:rep}
\langle\delta\phi_{\alpha}(\eta_{1},\textbf{x}_{1})\delta\phi_{\beta}(\eta_{2},\textbf{x}_{2})\rangle = \int \mathcal{D} \delta\phi\ \delta\phi_{\alpha}(\eta_{1},\textbf{x}_{1})\delta\phi_{\beta}(\eta_{2},\textbf{x}_{2})\ e^{i\mathcal{S}}
\end{equation}
using path integral formulation. Let \eqref{eq:cor} be dragged along the flow lines of vector field \(\xi\), such that it is an isometry of the space and the symmetry is non-anomalous. Hence,
\begin{equation}\label{eq:lie}
L_{\xi}\langle\delta\phi_{\alpha}(\eta_{1},\textbf{x}_{1})\delta\phi_{\beta}(\eta_{2},\textbf{x}_{2})\rangle = 0
\end{equation} 
Using \eqref{eq:lie} in \eqref{eq:rep}
\begin{equation}\label{eq:one}
\begin{split}
0 = \int \mathcal{D}\delta\phi\ & \Bigl[L_{\xi}\delta\phi_{\alpha}(\eta_{1},\textbf{x}_{1})\ \delta\phi_{\beta}(\eta_{2},\textbf{x}_{2}) + \delta\phi_{\alpha}(\eta_{1},\textbf{x}_{1})\ L_{\xi}\delta\phi_{\beta}(\eta_{2},\textbf{x}_{2})\\
                               & + i\ L_{\xi}\mathcal{S}\ \delta\phi_{\alpha}(\eta_{1},\textbf{x}_{1}) \delta\phi_{\beta}(\eta_{2},\textbf{x}_{2})
                                       \Bigr]\ e^{i\mathcal{S}}
\end{split}
\end{equation}
Substituting \eqref{eq:ac} in \eqref{eq:one} and evaluating the Lie derivative
\begin{equation}
\begin{split}
0 = \int \mathcal{D}\delta\phi\int d^{3}xd\eta &\ \Bigl[\delta(\textbf{x}-\textbf{x}_{1})\delta(\eta-\eta_{1})L_{\xi}\delta\phi_{\alpha}(\eta,\textbf{x})\delta\phi_{\beta}(\eta_{2},\textbf{x}_{2})\\
                                               & + \delta(\textbf{x}-\textbf{x}_{2})\delta(\eta-\eta_{2})\delta\phi_{\alpha}(\eta_{1},\textbf{x}_{1})L_{\xi}\delta\phi_{\beta}(\eta,\textbf{x})\\
                                               & + i\ L_{\xi}\mathcal{L}\ \delta\phi_{\alpha}(\eta_{1},\textbf{x}_{1})\delta\phi_{\beta}(\eta_{2},\textbf{x}_{2})\Bigr]\ e^{i\mathcal{S}}
\end{split}
\end{equation}
Rearranging the terms and writing it in the bracket notation,
\begin{equation}\label{eq:wi}
\begin{split}
\langle L_{\xi}\mathcal{L}\ \delta\phi_{\alpha}(\eta_{1},\textbf{x}_{1})\delta\phi_{\beta}(\eta_{2},\textbf{x}_{2}) \rangle = i &\ \Bigl[ \delta(\textbf{x}-\textbf{x}_{1})\delta(\eta-\eta_{1})\langle L_{\xi}\delta\phi_{\alpha}(\eta,\textbf{x})\delta\phi_{\beta}(\eta_{2},\textbf{x}_{2})\rangle \\
                                                                                                                                & + \delta(\textbf{x}-\textbf{x}_{2})\delta(\eta-\eta_{2})\langle\delta\phi_{\alpha}(\eta_{1},\textbf{x}_{1})L_{\xi}\delta\phi_{\beta}(\eta,\textbf{x}) \rangle \Bigr]
\end{split}
\end{equation}
Eqn \eqref{eq:wi} is the WI for two-point function. 
\section{Killing Vectors}\label{sec:vec}
The vector field $\xi$ in \eqref{eq:lie} with respect to which $C$ is differentiated is a Killing vector; generator of isometry for the Robertson-Walker space-time whose metric is given as
\begin{equation}\label{eq:met}
ds^{2} = a^{2}(\eta)\ (d\eta^{2} - dx^{2} - dy^{2} - dz^{2})
\end{equation}
where
\begin{equation}\label{eq:gmn}
g_{\mu\nu} = diag\ (a^{2}(\eta), -a^{2}(\eta), -a^{2}(\eta), -a^{2}(\eta))
\end{equation}
is the metric tensor. The vector field $\xi$ is determined by the Killing equation
\begin{equation}\label{eq:kill}
\nabla_{\mu}\ \xi_{\nu} + \nabla_{\nu}\ \xi_{\mu} = 0
\end{equation}
where $\nabla$ represents the covariant derivative and \eqref{eq:kill} is valid for contravariant components
\begin{equation}\label{eq:killcon}
\nabla_{\mu}\ \xi^{\nu} + \nabla_{\nu}\ \xi^{\mu} = 0\ .
\end{equation}
Eqns. \eqref{eq:kill}-\eqref{eq:killcon} are derived from the condition
\begin{equation}
L_{\xi}\ g_{\mu\nu} = \xi^{\sigma}\partial_{\sigma}g_{\mu\nu} + g_{\sigma\nu}\partial_{\mu}\xi^{\sigma} + g_{\mu\sigma}\partial_{\nu}\xi^{\sigma} = 0
\end{equation}
where compatibility of the metric
\begin{equation}
\nabla_{\sigma}g_{\mu\nu} = 0
\end{equation}
is understood. 
\subsection{Generators}
In order to evaluate \eqref{eq:kill} it is required to calculate the Christoffel symbols which is given by 
\begin{equation}
\Gamma^{\sigma}_{\mu\nu} = \frac{1}{2}\ g^{\sigma\gamma}\ (g_{\gamma\mu,\nu} + g_{\gamma\nu,\mu} - g_{\mu\nu,\gamma})\ .
\end{equation}
From \eqref{eq:gmn} the metric components are 
\begin{equation}\label{eq:comp}
g_{\eta\eta} = a^{2},\ g^{\eta\eta} = \frac{1}{a^{2}};\ g_{ij} = -a^{2}\delta_{ij},\ g^{ij} = -\frac{1}{a^{2}}\delta^{ij}
\end{equation}
and the Christoffel symbols are as follows:
\begin{equation}\label{eq:chr}
\Gamma^{\eta}_{\eta\eta} = \frac{a^{'}}{a},\ \Gamma^{i}_{j\eta} = \frac{a^{'}}{a}\ \delta^{i}_{j},\ \Gamma^{\eta}_{ij} = \frac{a^{'}}{a}\ \delta_{ij}
\end{equation}
and
\begin{equation}\label{eq:chrzer}
\Gamma^{i}_{ij} = \Gamma^{i}_{\eta\eta} = \Gamma^{\eta}_{\eta i} = 0
\end{equation}
From \eqref{eq:kill} 
\begin{equation}
g_{\nu\sigma}\nabla_{\mu}\xi^{\sigma} + g_{\mu\sigma}\nabla_{\nu}\xi^{\sigma} = 0
\end{equation}
and expanding the covariant derivative we get
\begin{equation}\label{eq:eval}
g_{\nu\sigma}(\partial_{\mu}\xi^{\sigma} + \Gamma^{\sigma}_{\mu\gamma}\xi^{\gamma}) + g_{\mu\sigma}(\partial_{\nu}\xi^{\sigma} + \Gamma^{\sigma}_{\nu\gamma}\xi^{\gamma}) = 0 \ .
\end{equation}
\begin{table}[h]\label{tab:gen}
\centering
\caption[Generators]{Generators}
\begin{tabular}{| c | c | c |}
\hline
Symmetry & Symbol & Killing Vector\ ($\xi^{\mu}\partial_{\mu}$)  \\ \hline
1-Temporal & $T$ & $\frac{1}{a}\partial_{\eta}$\\ \hline
3-Spatial & $S$ & $\partial_{i}$\\ \hline
3-Lorentz Boosts & $L$ & $x^{i}\partial_{\eta} + \eta\partial_{i}$\\ \hline
3-Rotation & $J$ & $-i\ (x^{i}\partial_{j} - x^{j}\partial_{i})$ \\
\hline
\end{tabular}
\end{table}
Making appropriate choices for the indices $\mu$ and $\nu$ , in terms of temporal and spatial coordinates and also using the fact that the metric is symmetric with no off-diagonal terms we arrive at 10 linearly independent Killing vectors.
\subsection{Commutation Relations}
The set of Killing vectors which preserves the metric, hence generators of isometry, are the elements forming a Lie group. From \ref{tab:gen}, the commutation relations among the generators of isometry forms the associated Lie algebra given in \ref{tab:lie}. It is to be noted that commutation relation for any two Killing vectors from \ref{tab:gen}, results in a Killing vector which belongs to the same set as shown in \ref{tab:lie}.
\begin{sidewaystable}[h]\label{tab:lie}
\centering
\caption[Lie Algebra]{Lie Algebra}
\begin{tabular}{|c|c|c|c|c|c|c|c|c|c|c|}
\hline
$[.,.]$ & $T$ & $S_{x}$ & $S_{y}$ & $S_{z}$ & $L_{x}$ & $L_{y}$ & $L_{z}$ & $J_{x}$ & $J_{y}$ & $J_{z}$\\ \hline
$T$ & $0$ & $0$ & $0$ & $0$ & $\frac{1}{a}S_{x}+\mathcal{H}x\ T$ & $\frac{1}{a}S_{y}+\mathcal{H}y\ T$ & $\frac{1}{a}S_{z}+\mathcal{H}z\ T$ & $0$ & $0$ & $0$\\ \hline
$S_{x}$ & $0$ & $0$ & $0$ & $0$ & $aT$ & $0$ & $0$ & $0$ & $iS_{z}$ & $-iS_{y}$\\ \hline
$S_{y}$ & $0$ & $0$ & $0$ & $0$ & $0$ & $aT$ & $0$ & $-iS_{z}$ & $0$ & $iS_{x}$\\ \hline
$S_{z}$ & $0$ & $0$ & $0$ & $0$ & $0$ & $0$ & $aT$ & $iS_{y}$ & $-iS_{x}$ & $0$\\ \hline
$L_{x}$ & $- (\frac{1}{a}S_{x}+\mathcal{H}x\ T)$ & $-aT$ & $0$ & $0$ & $0$ & $iJ_{z}$ & $-iJ_{y}$ & $0$ & $iL_{z}$ & $-iL_{y}$\\ \hline
$L_{y}$ & $- (\frac{1}{a}S_{y}+\mathcal{H}y\ T)$ & $0$ & $-aT$ & $0$ & $-iJ_{z}$ & $0$ & $iJ_{x}$ & $-iL_{z}$ & $0$ & $iL_{x}$\\ \hline
$L_{z}$ & $- (\frac{1}{a}S_{z}+\mathcal{H}z\ T)$ & $0$ & $0$ & $-aT$ & $iJ_{y}$ & $-iJ_{x}$ & $0$ & $iL_{y}$ & $-iL_{x}$ & $0$\\ \hline
$J_{x}$ & $0$ & $0$ & $iS_{z}$ & $-iS_{y}$ & $0$ & $iL_{z}$ & $-iL_{y}$ & $0$ & $iJ_{z}$ & $-iJ_{y}$\\ \hline
$J_{y}$ & $0$ & $-iS_{z}$ & $0$ & $iS_{x}$ & $-iL_{z}$ & $0$ & $iL_{x}$ & $-iJ_{z}$ & $0$ & $iJ_{x}$\\ \hline
$J_{z}$ & $0$ & $iS_{y}$ & $-iS_{x}$ & $0$ & $iL_{y}$ & $-iL_{x}$ & $0$ & $iJ_{y}$ & $-iJ_{x}$ & $0$\\
\hline
\end{tabular}
\end{sidewaystable}
\section{Algebra of Ward Identity}
The WI was derived in \eqref{eq:wi} considering the Lie derivative of two-point function defined in \eqref{eq:cor}, with symmetry being non-anomalous. We now extend this derivation to a more generic analysis and prove an important result that poses an algebraic constraint to the WI in terms of commutation relations.
\subsection{Proof}
Consider the two-point function,
\begin{equation}
C((\eta_{1},\textbf{x}_{1}),(\eta_{2},\textbf{x}_{2})) := \langle\delta\phi_{\alpha}(\eta_{1},\textbf{x}_{1})\delta\phi_{\beta}(\eta_{2},\textbf{x}_{2})\rangle
\end{equation}
and take its Lie derivative
\begin{equation}\label{eq:prof}
L_{\xi}\langle\delta\phi_{\alpha}(\eta_{1},\textbf{x}_{1})\delta\phi_{\beta}(\eta_{2},\textbf{x}_{2})\rangle = 0
\end{equation} 
It is trivial to know that taking Lie derivative of \eqref{eq:prof} is still $0$, such that
\begin{equation}\label{eq:ana}
L_{\xi}L_{\xi}\langle\delta\phi_{\alpha}(\eta_{1},\textbf{x}_{1})\delta\phi_{\beta}(\eta_{2},\textbf{x}_{2})\rangle = 0
\end{equation}
From the definition of Lie derivative for scalar fields it is know that
\begin{equation}
L_{\xi}\phi(x) = \xi^{\mu}\partial_{\mu}\phi(x)
\end{equation}
where $\xi^{\mu}$ are the components of a vector field. Apply this definition to \eqref{eq:ana}, where $\xi$'s are Killing vectors and choose two distinct Killing vectors, i.e. dragging them along distinct flow lines of the vector field, followed by reversing the order of Lie differentiation and subtracting them, we get
\begin{equation}
(L_{\xi^{a}}L_{\xi^{b}} - L_{\xi^{b}}L_{\xi^{a}})\ \langle\delta\phi_{\alpha}(\eta_{1},\textbf{x}_{1})\delta\phi_{\beta}(\eta_{2},\textbf{x}_{2})\rangle = 0
\end{equation}
hence,
\begin{equation}\label{eq:comm}
[L_{\xi^{a}},L_{\xi^{b}}] \langle\delta\phi_{\alpha}(\eta_{1},\textbf{x}_{1})\delta\phi_{\beta}(\eta_{2},\textbf{x}_{2})\rangle = 0
\end{equation}
where $\xi^{a}$ and $\xi^{b}$ are Killing vectors from \ref{tab:gen}. The \textit{closure property} in \eqref{eq:comm} as obtained from \ref{tab:lie} is manifest in the WI \eqref{eq:wi}.
\paragraph{}
It is to be noted that the derivation of WI was done by taking Lie derivative of the two-point function. In this analysis the first step corresponds to 
\begin{equation}
L_{\xi^{b}} \langle\delta\phi_{\alpha}(\eta_{1},\textbf{x}_{1})\delta\phi_{\beta}(\eta_{2},\textbf{x}_{2})\rangle = 0
\end{equation}
which leads to 
\begin{equation}\label{eq:wio}
\begin{split}
\langle L_{\xi^{b}}\mathcal{L}\ \delta\phi_{\alpha}(\eta_{1},\textbf{x}_{1})\delta\phi_{\beta}(\eta_{2},\textbf{x}_{2}) \rangle = i &\ \Bigl[ \delta(\textbf{x}-\textbf{x}_{1})\delta(\eta-\eta_{1})\langle L_{\xi^{b}}\delta\phi_{\alpha}(\eta,\textbf{x})\delta\phi_{\beta}(\eta_{2},\textbf{x}_{2})\rangle \\
                                                                                                                                & + \delta(\textbf{x}-\textbf{x}_{2})\delta(\eta-\eta_{2})\langle\delta\phi_{\alpha}(\eta_{1},\textbf{x}_{1})L_{\xi^{b}}\delta\phi_{\beta}(\eta,\textbf{x}) \rangle \Bigr]
\end{split}
\end{equation}
the WI. Applying another Lie derivative to \eqref{eq:wio}
\begin{equation}\label{eq:wit}
\begin{split}
L_{\xi^{a}}\langle L_{\xi^{b}}\mathcal{L}\ \delta\phi_{\alpha}(\eta_{1},\textbf{x}_{1})\delta\phi_{\beta}(\eta_{2},\textbf{x}_{2}) \rangle = i &\ \Bigl[ \delta(\textbf{x}-\textbf{x}_{1})\delta(\eta-\eta_{1})L_{\xi^{a}}\langle L_{\xi^{b}}\delta\phi_{\alpha}(\eta,\textbf{x})\delta\phi_{\beta}(\eta_{2},\textbf{x}_{2})\rangle \\
                                                                                                                                & + \delta(\textbf{x}-\textbf{x}_{2})\delta(\eta-\eta_{2})L_{\xi^{a}}\langle\delta\phi_{\alpha}(\eta_{1},\textbf{x}_{1})L_{\xi^{b}}\delta\phi_{\beta}(\eta,\textbf{x}) \rangle \Bigr]
\end{split}
\end{equation}
followed by reversing the order of differentiation, we get
\begin{equation}\label{eq:wi3}
\begin{split}
L_{\xi^{b}}\langle L_{\xi^{a}}\mathcal{L}\ \delta\phi_{\alpha}(\eta_{1},\textbf{x}_{1})\delta\phi_{\beta}(\eta_{2},\textbf{x}_{2}) \rangle = i &\ \Bigl[ \delta(\textbf{x}-\textbf{x}_{1})\delta(\eta-\eta_{1})L_{\xi^{b}}\langle L_{\xi^{a}}\delta\phi_{\alpha}(\eta,\textbf{x})\delta\phi_{\beta}(\eta_{2},\textbf{x}_{2})\rangle \\
                                                                                                                                & + \delta(\textbf{x}-\textbf{x}_{2})\delta(\eta-\eta_{2})L_{\xi^{b}}\langle\delta\phi_{\alpha}(\eta_{1},\textbf{x}_{1})L_{\xi^{a}}\delta\phi_{\beta}(\eta,\textbf{x}) \rangle \Bigr].
\end{split}
\end{equation}
Eqns.\eqref{eq:wit}-\eqref{eq:wi3} are evaluated by the path integral method. As it is beyond scope to show the explicit calculations which where carried out, the general method is explained. On taking the Lie derivative for the second time, it acts on every piece present within $\langle.\rangle$ on both sides, which also includes the action and hence the Lagrangian density from the term $e^{i\mathcal{S}}$, present in the definition of path integral representation for two-point function. After evaluation, the expressions obtained from Eqns.\eqref{eq:wit}-\eqref{eq:wi3} are subtracted and survival terms are written in the bracket-notation, involving the commutation relation 
\begin{equation}\label{eq:wif}
\begin{split}
\langle [L_{\xi^{a}},L_{\xi^{b}}]\mathcal{L}\ \delta\phi_{\alpha}(\eta_{1},\textbf{x}_{1})\delta\phi_{\beta}(\eta_{2},\textbf{x}_{2}) \rangle = i &\ \Bigl[ \delta(\textbf{x}-\textbf{x}_{1})\delta(\eta-\eta_{1})\langle [L_{\xi^{a}}, L_{\xi^{b}}]\delta\phi_{\alpha}(\eta,\textbf{x})\delta\phi_{\beta}(\eta_{2},\textbf{x}_{2})\rangle \\
                                                                                                                                & + \delta(\textbf{x}-\textbf{x}_{2})\delta(\eta-\eta_{2})\langle\delta\phi_{\alpha}(\eta_{1},\textbf{x}_{1})[L_{\xi^{a}},L_{\xi^{b}}]\delta\phi_{\beta}(\eta,\textbf{x}) \rangle \Bigr]
\end{split}
\end{equation}
which establishes the \textit{closure property} of WI as follows; the commutator bracket in \eqref{eq:wif} follows the algebra as shown in \ref{tab:lie} corresponding to the choice of Killing vectors shown in \ref{tab:gen}, which is manifested in \eqref{eq:comm}. For example, if $\xi^{a}$ corresponds to rotation in x-direction $(J_{x})$ and $\xi^{b}$ to rotation in y-direction $(J_{y})$ then from \eqref{eq:comm} and \eqref{eq:wif} we get the WI corresponding to rotation in z-direction $(J_{z})$ and the structure constant for the identity is $1$ or $-1$ depending on the order of commutation. This result holds for all commutation relations that the 10 linearly independent Killing vectors obey, corresponding structure constants being the permutation tensor $\epsilon_{abc}$, which picks up the value $1$ or $-1$ depending on cyclic or anti-cyclic permutation which reflects the order of commutation. Therefore, WI itself follows a Lie algebra, such that
\begin{equation}\label{eq:res}
[ L_{\xi^{a}}, L_{\xi^{b}} ]\ C((\eta_{1},\textbf{x}_{1}),(\eta_{2},\textbf{x}_{2})) = \epsilon_{abc}L_{\xi^{c}}\ C((\eta_{1},\textbf{x}_{1}),(\eta_{2},\textbf{x}_{2}))
\end{equation}
hence,
\begin{equation}\label{eq:resf}
\begin{split}
\epsilon_{abc}\langle L_{\xi^{c}}\mathcal{L}\ \delta\phi_{\alpha}(\eta_{1},\textbf{x}_{1})\delta\phi_{\beta}(\eta_{2},\textbf{x}_{2}) \rangle = i \epsilon_{abc} &\ \Bigl[ \delta(\textbf{x}-\textbf{x}_{1})\delta(\eta-\eta_{1})\langle L_{\xi^{c}}\delta\phi_{\alpha}(\eta,\textbf{x})\delta\phi_{\beta}(\eta_{2},\textbf{x}_{2})\rangle \\
                                                                                                                                & + \delta(\textbf{x}-\textbf{x}_{2})\delta(\eta-\eta_{2})\langle\delta\phi_{\alpha}(\eta_{1},\textbf{x}_{1})L_{\xi^{c}}\delta\phi_{\beta}(\eta,\textbf{x}) \rangle \Bigr]
\end{split}
\end{equation}
\textbf{QED}
\section{Equation for Conservation}
The WI as obtained in \eqref{eq:wi} is now solved by evaluating the Lie derivative of the Lagrangian density $\mathcal{L}$ as follows
\begin{equation}\label{eq:der}
L_{\xi}\mathcal{L} = \xi^{\mu}\partial_{\mu}\mathcal{L}
\end{equation}
where
\begin{equation}\label{eq:lag}
\partial_{\mu}\mathcal{L} = \frac{\partial\mathcal{L}}{\partial(\partial_{\nu}\delta\phi_{\lambda})}\frac{\partial(\partial_{\nu}\delta\phi_{\lambda})}{\partial x^{\mu}} + \frac{\partial\mathcal{L}}{\partial(\delta\phi_{\lambda})}\frac{\partial(\delta\phi_{\lambda})}{\partial x^{\mu}} \ .
\end{equation}
From Euler-Lagrange equation of motion,
\begin{equation}\label{eq:eul}
\partial_{\nu}\left(\frac{\partial\mathcal{L}}{\partial(\partial_{\nu}\delta\phi_{\lambda})}\right) = \frac{\partial\mathcal{L}}{\partial(\delta\phi_{\lambda})} 
\end{equation}
Substituting \eqref{eq:eul} in \eqref{eq:lag} 
\begin{equation}\label{eq:ans}
\partial_{\mu}\mathcal{L}  = \partial_{\nu}\left(\frac{\partial\mathcal{L}}{\partial(\partial_{\nu}\delta\phi_{\lambda})}\partial_{\mu}\delta\phi_{\lambda}\right) \ .
\end{equation}
The energy-momentum tensor is defined as
\begin{equation}\label{eq:emt}
T^{\nu}\ _{\mu} := \frac{\partial\mathcal{L}}{\partial(\partial_{\nu}\delta\phi_{\lambda})}\partial_{\mu}\delta\phi_{\lambda} - \mathcal{L}\ \delta^{\nu}\ _{\mu} 
\end{equation}
Substituting \eqref{eq:der}, \eqref{eq:ans} and \eqref{eq:emt} in \eqref{eq:wi} followed by cancelling off terms on both sides, we get
\begin{equation}
\xi^{\nu}\langle \partial_{\mu}T^{\mu}\ _{\nu}\ \delta\phi_{\alpha}(\eta_{1},\textbf{x}_{1})\delta\phi_{\beta}(\eta,\textbf{x}_{2})\rangle = 0
\end{equation}
where $\xi^{\nu}$ are components of Killing vectors and in case of model independent analysis they are arbitrary components of a vector field, hence
\begin{equation}\label{eq:cons}
\langle \partial_{\mu}T^{\mu}\ _{\nu}\ \delta\phi_{\alpha}(\eta_{1},\textbf{x}_{1})\delta\phi_{\beta}(\eta,\textbf{x}_{2})\rangle = 0 \ .
\end{equation}
which is the conservation equation for the multi-field Lagrangian density considered in \eqref{eq:inf}. 
\subsection{D'Alembertian Relation}
The components of energy-momentum tensor for Lagrangian density \eqref{eq:inf} as obtained from \eqref{eq:emt} are:
\begin{equation}
T^{\eta}\ _{\nu} = \frac{1}{2} a^{2}\partial^{\eta}\delta\phi_{\lambda}(\eta,\textbf{x})\partial_{\nu}\delta\phi_{\lambda}(\eta,\textbf{x}),\ T^{i}\ _{\nu} = -\frac{1}{2} a^{2}\partial^{i}\delta\phi_{\lambda}(\eta,\textbf{x})\partial_{\nu}\delta\phi_{\lambda}(\eta,\textbf{x})
\end{equation}
Making use of the above result in \eqref{eq:cons}
\begin{equation}
\langle \partial_{\eta}T^{\eta}\ _{\nu}\delta\phi_{\alpha}(\eta_{1},\textbf{x}_{1})\ \delta\phi_{\beta}(\eta_{2},\textbf{x}_{2})\rangle + \langle \partial_{i}T^{i}\ _{\nu}\delta\phi_{\alpha}(\eta_{1},\textbf{x}_{1})\ \delta\phi_{\beta}(\eta_{2},\textbf{x}_{2})\rangle = 0
\end{equation}
Substituting the components of energy-momentum tensor as obtained and solving the above equation, we get
\begin{equation}
\begin{split}
\langle \partial_{\eta}(a^{2}\partial^{\eta}\delta\phi_{\lambda}(\eta,\textbf{x}) & \partial_{\nu}\delta\phi_{\lambda}(\eta,\textbf{x}))\delta\phi_{\alpha}(\eta_{1},\textbf{x}_{1})\delta\phi_{\beta}(\eta_{1},\textbf{x}_{2})\rangle \\
                                                                                  & - \langle a^{2} \partial_{i}(\partial^{i}\delta\phi_{\lambda}(\eta,\textbf{x})\partial_{\nu}\delta\phi_{\lambda}(\eta,\textbf{x}))\delta\phi_{\alpha}(\eta_{1},\textbf{x}_{1})\delta\phi_{\beta}(\eta_{1},\textbf{x}_{2})\rangle = 0
\end{split}
\end{equation}

\begin{equation}
\begin{split}
2aa^{'}\partial^{\eta}\partial_{\nu}\langle\delta\phi_{\lambda} & (\eta,\textbf{x})\delta\phi_{\lambda}(\eta,\textbf{x})\delta\phi_{\alpha}(\eta_{1},\textbf{x}_{1})\delta\phi_{\beta}(\eta_{1},\textbf{x}_{2})\rangle \\
                                                                & a^{2}\square\partial_{\nu}\langle\delta\phi_{\lambda}(\eta,\textbf{x})\delta\phi_{\lambda}(\eta,\textbf{x})\delta\phi_{\alpha}(\eta_{1},\textbf{x}_{1})\delta\phi_{\beta}(\eta_{1},\textbf{x}_{2})\rangle = 0 \ .
\end{split}
\end{equation}
Dividing throughout by $a^{2}$
\begin{equation}\label{eq:box}
\begin{split}
\square \partial_{\nu}\langle\delta\phi_{\lambda}(\eta,\textbf{x}) & \delta\phi_{\lambda}(\eta,\textbf{x})\delta\phi_{\alpha}(\eta_{1},\textbf{x}_{1})\delta\phi_{\beta}(\eta_{1},\textbf{x}_{2})\rangle \\
								& = -2\mathcal{H}\partial^{\eta}\partial_{\nu}\langle\delta\phi_{\lambda}(\eta,\textbf{x})\delta\phi_{\lambda}(\eta,\textbf{x})\delta\phi_{\alpha}(\eta_{1},\textbf{x}_{1})\delta\phi_{\beta}(\eta_{1},\textbf{x}_{2})\rangle
\end{split}
\end{equation}
This relation will aid in arriving at an equation of motion.
\section{Equation of Motion}
Consider 
\begin{equation}\label{eq:one}
\langle \partial_{\mu}T^{\mu}\ _{\nu}\ \delta\phi_{\alpha}(\eta_{1},\textbf{x}_{1})\delta\phi_{\beta}(\eta,\textbf{x}_{2})\rangle = 0 
\end{equation}
where 
\begin{equation}\label{eq:two}
T^{\mu}\ _{\nu} = \frac{\partial \mathcal{L}}{\partial(\partial_{\mu}\delta\phi_{\lambda})}\partial_{\nu}(\delta\phi_{\lambda}) - \mathcal{L}\ \delta^{\mu}\ _{\nu}
\end{equation}
Substituting \eqref{eq:two} in \eqref{eq:one} 
\begin{equation}
\begin{split}
0 =  \left<\partial_{\mu}\left(\frac{\partial \mathcal{L}}{\partial(\partial_{\mu}\delta\phi_{\lambda})}\partial_{\nu}\delta\phi_{\lambda}\right)\delta\phi_{\alpha}(\eta_{1},\textbf{x}_{1})\delta\phi_{\beta}(\eta_{2},\textbf{x}_{2})\right>  - \\
                 & \left<\partial_{\nu}\mathcal{L}\ \delta\phi_{\alpha}(\eta_{1},\textbf{x}_{1})\delta\phi_{\beta}(\eta_{2},\textbf{x}_{2})\right> 
\end{split}
\end{equation}
The term present within the parentheses of the first object in the above equation is solved in order to get
\begin{equation}
\begin{split}
0 =  2\partial_{\mu}\partial_{\nu}\left<\left(\frac{\partial \mathcal{L}}{\partial(\partial_{\mu}\delta\phi_{\lambda})}\delta\phi_{\lambda}\right)\delta\phi_{\alpha}(\eta_{1},\textbf{x}_{1})\delta\phi_{\beta}(\eta_{2},\textbf{x}_{2})\right>  - \\
                 & \left<\partial_{\nu}\mathcal{L}\ \delta\phi_{\alpha}(\eta_{1},\textbf{x}_{1})\delta\phi_{\beta}(\eta_{2},\textbf{x}_{2})\right> .
\end{split}
\end{equation}
First object is solved by applying summation over $\mu$ and making appropriate substitution for $\left(\frac{\partial \mathcal{L}}{\partial(\partial_{\mu}\delta\phi_{\lambda})}\partial_{\nu}\delta\phi_{\lambda}\right)$ we get
\begin{equation}
\begin{split}
\partial_{\nu}\partial_{\eta}\langle a^{2}\partial^{\eta}\delta\phi_{\lambda} & (\eta,\textbf{x})\delta\phi_{\lambda}(\eta,\textbf{x})\delta\phi_{\alpha}(\eta_{1},\textbf{x}_{1})\delta\phi_{\beta}(\eta_{2},\textbf{x}_{2})\rangle \\
									                                          & -\partial_{\nu}\partial_{i}\langle a^{2}\partial^{i}\delta\phi_{\lambda}(\eta,\textbf{x})\delta\phi_{\lambda}(\eta,\textbf{x})\delta\phi_{\alpha}(\eta_{1},\textbf{x}_{1})\delta\phi_{\beta}(\eta_{2},\textbf{x}_{2})\rangle
\end{split}
\end{equation}
which on solving leads to 
\begin{equation}
\begin{split}
2a^{2}\mathcal{H}\delta^{\eta}\ _{\nu} \square \langle \delta\phi_{\lambda}\delta\phi_{\lambda}\delta\phi_{\alpha}\delta_{\beta}\rangle + 2\partial_{\nu}(a^{2}\mathcal{H})\partial^{\eta}\langle \delta\phi_{\lambda}\delta \phi_{\lambda}\delta\phi_{\alpha}\delta_{\beta}\rangle + 2a^{2}\mathcal{H}\partial^{\eta}\partial_{\nu} & \langle \delta\phi_{\lambda}\delta\phi_{\lambda}\delta\phi_{\alpha}\delta_{\beta}\rangle \\
               & a^{2}\square \partial_{\nu}\langle \delta\phi_{\lambda}\delta\phi_{\lambda}\delta\phi_{\alpha}\delta_{\beta}\rangle
\end{split}
\end{equation}
Making use of \eqref{eq:box}, cancels the last two objects and finally gives
\begin{equation}\label{eq:ek}
2a^{2}\mathcal{H}\delta^{\eta}\ _{\nu} \square \langle \delta\phi_{\lambda}\delta\phi_{\lambda}\delta\phi_{\alpha}\delta_{\beta}\rangle + 2\partial_{\nu}(a^{2}\mathcal{H})\partial^{\eta}\langle \delta\phi_{\lambda}\delta \phi_{\lambda}\delta\phi_{\alpha}\delta_{\beta}\rangle
\end{equation}
It is now required to evaluate $\partial_{\nu}\mathcal{L}$ which contains the mass and potential terms given as
\begin{equation}\label{eq:do}
\begin{split}
\frac{1}{2!}m_{\kappa\omega}\partial_{\nu}(a^{4})\langle\delta\phi_{\kappa}\delta\phi_{\omega}\delta\phi_{\alpha}\delta\phi_{\beta}\rangle+\frac{1}{2!}m_{\kappa\omega}(a^{4})\partial_{\nu}\langle\delta\phi_{\kappa}&\delta\phi_{\omega}\delta\phi_{\alpha}\delta\phi_{\beta}\rangle \\
        &+\frac{1}{3!}V_{\kappa\omega\zeta}\partial_{\nu}(a^{4})\langle\delta\phi_{\kappa}\delta\phi_{\omega}\delta\phi_{\zeta}\delta\phi_{\alpha}\delta\phi_{\beta}\rangle \\
        &+\frac{1}{3!}V_{\kappa\omega\zeta}(a^{4})\partial_{\nu}\langle\delta\phi_{\kappa}\delta\phi_{\omega}\delta\phi_{\zeta}\delta\phi_{\alpha}\delta\phi_{\beta}\rangle
\end{split}
\end{equation}
Finally adding up \eqref{eq:ek} and \eqref{eq:do}
\begin{equation}\label{eq:eom}
\begin{split}
2a^{2}&\mathcal{H}\delta^{\eta}\ _{\nu} \square \langle \delta\phi_{\lambda}\delta\phi_{\lambda}\delta\phi_{\alpha}\delta_{\beta}\rangle + \partial_{\nu}(a^{2}\mathcal{H})\partial^{\eta}\langle\delta\phi_{\lambda}\delta \phi_{\lambda}\delta\phi_{\alpha}\delta_{\beta}\rangle+\frac{1}{2!}m_{\kappa\omega}\partial_{\nu}(a^{4})\langle\delta\phi_{\kappa}\delta\phi_{\omega}\delta\phi_{\alpha}\delta\phi_{\beta}\rangle \\+
      &\frac{1}{2!}m_{\kappa\omega}(a^{4})\partial_{\nu}\langle\delta\phi_{\kappa}\delta\phi_{\omega}\delta\phi_{\alpha}\delta\phi_{\beta}\rangle+\frac{1}{3!}V_{\kappa\omega\zeta}\partial_{\nu}(a^{4})\langle\delta\phi_{\kappa}\delta\phi_{\omega}\delta\phi_{\zeta}\delta\phi_{\alpha}\delta\phi_{\beta}\rangle + \\
      &  \frac{1}{3!}V_{\kappa\omega\zeta}(a^{4})\partial_{\nu}\langle\delta\phi_{\kappa}\delta\phi_{\omega}\delta\phi_{\zeta}\delta\phi_{\alpha}\delta\phi_{\beta}\rangle = 0       
\end{split}
\end{equation}
This is the equation of motion, as obtained by solving \eqref{eq:cons} which in turn was derived from the WI for two-point function as obtained in \eqref{eq:wi}. 
\paragraph{}
There are two cases for the equation of motion, one corresponding to $\nu = \eta$ and another being $\nu \neq \eta$. When $\nu = \eta$
\begin{equation}\label{eq:time}
\begin{split}
2a^{2}\mathcal{H}&\square \langle \delta\phi_{\lambda}\delta\phi_{\lambda}\delta\phi_{\alpha}\delta_{\beta}\rangle+\mathcal{H}\ \Bigl(2\mathcal{H}+\frac{\mathcal{H}'}{\mathcal{H}}\Bigr)\partial_{\eta}\langle\delta\phi_{\lambda}\delta \phi_{\lambda}\delta\phi_{\alpha}\delta_{\beta}\rangle + \\
                 & \frac{1}{2!}m_{\kappa\omega}\partial_{\eta}(a^{4})\langle\delta\phi_{\kappa}\delta\phi_{\omega}\delta\phi_{\alpha}\delta\phi_{\beta}\rangle + \frac{1}{2!}m_{\kappa\omega}(a^{4})\partial_{\eta}\langle\delta\phi_{\kappa}\delta\phi_{\omega}\delta\phi_{\alpha}\delta\phi_{\beta}\rangle + \\
                 & \frac{1}{3!}V_{\kappa\omega\zeta}\partial_{\eta}(a^{4})\langle\delta\phi_{\kappa}\delta\phi_{\omega}\delta\phi_{\zeta}\delta\phi_{\alpha}\delta\phi_{\beta}\rangle \\
                 & + \frac{1}{3!}V_{\kappa\omega\zeta}(a^{4})\partial_{\eta}\langle\delta\phi_{\kappa}\delta\phi_{\omega}\delta\phi_{\zeta}\delta\phi_{\alpha}\delta\phi_{\beta}\rangle = 0       
\end{split}
\end{equation}
and for $\nu \neq \eta$
\begin{equation}\label{eq:space}
\frac{1}{2!}m_{\kappa\omega}\frac{\partial}{\partial \textbf{x}}\langle\delta\phi_{\kappa}\delta\phi_{\omega}\delta\phi_{\alpha}\delta\phi_{\beta}\rangle + \frac{1}{3!}V_{\kappa\omega\zeta}\frac{\partial}{\partial \textbf{x}}\langle\delta\phi_{\kappa}\delta\phi_{\omega}\delta\phi_{\zeta}\delta\phi_{\alpha}\delta\phi_{\beta}\rangle = 0 \ .
\end{equation}
\newpage
\section{Geometrical Approach}
From the equation for conservation
\begin{equation}
\langle \partial_{\mu}T^{\mu}\ _{\nu}\ \delta\phi_{\alpha}(\eta_{1},\textbf{x}_{1})\delta\phi_{\beta}(\eta,\textbf{x}_{2})\rangle = 0
\end{equation}
where the fields obtained from the energy-momentum tensor defined in \eqref{eq:emt} depend on $(\eta.\textbf{x})$, so up to the level of \textit{contact terms}
\begin{equation}\label{eq:sol}
\langle T^{\mu}\ _{\nu}\ \delta\phi_{\alpha}(\eta_{1},\textbf{x}_{1})\delta\phi_{\beta}(\eta,\textbf{x}_{2})\rangle = c^{\mu}\ _{\nu}
\end{equation}
hence,
\begin{equation}\label{eq:grav}
\langle G^{\mu}\ _{\nu}\ \delta\phi_{\alpha}(\eta_{1},\textbf{x}_{1})\delta\phi_{\beta}(\eta,\textbf{x}_{2})\rangle = 8 \pi G \ c^{\mu}\ _{\nu}
\end{equation}
where the object in RHS is a c-number. 
\subsection{Curvature Quantities}
Consider the metric defined in \eqref{eq:gmn}. The Riemann ( or curvature ) tensor is given as
\begin{equation}\label{eq:cur}
R^{\alpha}\ _{\beta\mu\nu} = \Gamma^{\alpha}_{\nu\beta,\mu}-\Gamma^{\alpha}_{\mu\beta,\nu}+\Gamma^{\delta}_{\nu\beta}\Gamma^{\alpha}_{\mu\delta}-\Gamma^{\delta}_{\mu\beta}\Gamma^{\alpha}_{\nu\delta}
\end{equation}
Using the result in \eqref{eq:chr} and exploiting the fact that metric is symmetric with no off-diagonal terms, we can solve \eqref{eq:cur} by making appropriate substitutions such that 
\begin{equation}
R^{\eta}\ _{i\eta j} = \partial_{\eta}\left(\frac{a'}{a}\right) \delta_{ij}
\end{equation}
\begin{equation}
R^{i}\ _{\eta j \eta} = - R^{\eta}\ _{i \eta j}\  \delta_{ij}
\end{equation}
\begin{equation}
R^{i}\ _{jij} = \left(\frac{a'}{a}\right)^{2}
\end{equation}
and contracting the Riemann tensor on first and third indices gives the Ricci tensor
\begin{equation}
R_{\eta\eta} = R^{i}\ _{\eta i \eta} = -3\ \partial_{\eta}\left(\frac{a'}{a}\right)
\end{equation}
\begin{equation}
R_{ij} = R^{\eta}\ _{i \eta j} + R^{k}\ _{ikj} = \left[\partial_{\eta}\left(\frac{a'}{a}\right)+2\ \left(\frac{a'}{a}\right)^{2}\right]\ \delta_{ij}
\end{equation} 
\newpage
which can be written in the form a matrix as

\[\begin{pmatrix}
-3\ \partial_{\eta}\left(\frac{a'}{a}\right) & 0 & 0 & 0\\
0 & \partial_{\eta}\left(\frac{a'}{a}\right)+2\ \left(\frac{a'}{a}\right)^{2} & 0 \\
0 & 0 & \partial_{\eta}\left(\frac{a'}{a}\right)+2\ \left(\frac{a'}{a}\right)^{2} & 0 \\
0 & 0 & 0 & \partial_{\eta}\left(\frac{a'}{a}\right)+2\ \left(\frac{a'}{a}\right)^{2}
\end{pmatrix}\]
The Ricci scalar is given by
\begin{equation}\label{eq:sca}
R \equiv g^{\mu\nu}R_{\mu\nu} = -\frac{6}{a^{2}}\left[\partial_{\eta}\left(\frac{a'}{a}\right)+\left(\frac{a'}{a}\right)^{2}\right] \ .
\end{equation}
As $R<0$ corresponding to space-time described by Robertson-Walker metric with 10 linearly independent Killing vectors as generators of isometry, the case is maximally symmetric.
\section{Probability Amplitude of Two-Point Function}
For a maximally symmetric space-time Ricci tensor and energy momentum tensor is proportional to the metric. Hence Einstein field equation is expressed as 
\begin{equation}\label{eq:ein}
G_{\mu\nu} \equiv R_{\mu\nu}-\frac{1}{2}g_{\mu\nu}\ R = \Lambda g_{\mu\nu}
\end{equation}
where energy-momentum tensor $T_{\mu\nu}$ is related to cosmological constant. 
We now relate the cosmological constant and normalised Ricci scalar $\kappa$ by setting $\mu = \nu = \eta$ in \eqref{eq:ein}
\begin{equation}
R_{\eta\eta} - \frac{1}{2}g_{\eta\eta}R = \Lambda \ g_{\eta\eta}
\end{equation}
which on solving gives
\begin{equation}
\left(\frac{a'}{a}\right) = \pm \sqrt{\frac{\Lambda}{3}} \ a
\end{equation}
Substituting this in \eqref{eq:sca} we get
\begin{equation}
R = - 4 \ \Lambda \ .
\end{equation}
As 
\begin{equation}
\kappa = \frac{R}{n(n-1)}
\end{equation}
where $n=4$, 
\begin{equation}
\Lambda = -3\kappa \ .
\end{equation}
Hence,
\begin{equation}\label{eq:reso}
T_{\mu\nu} = -\frac{3\kappa}{8\pi G}g_{\mu\nu}
\end{equation}
where $G_{\mu\nu} = 8\pi G\ T_{\mu\nu}$. From \eqref{eq:sol}
\begin{equation}
g^{\mu\rho}\langle T_{\rho\nu}\delta\phi_{\alpha}(\eta_{1},\textbf{x}_{1})\delta\phi_{\beta}(\eta,\textbf{x}_{2}) \rangle = c^{\mu}\ _{\nu}
\end{equation}
Substituting \eqref{eq:reso} in the above equation
\begin{equation}
\langle\delta\phi_{\alpha}(\eta_{1},\textbf{x}_{1})\delta\phi_{\beta}(\eta,\textbf{x}_{2})\rangle = \frac{8\pi G}{\Lambda}\ \delta^{\nu}\ _{\mu}c^{\mu}\ _{\nu}
\end{equation}
where $\rho_{\Lambda} = \frac{\Lambda}{8\pi G}$. Therefore,
\begin{equation}\label{eq:final}
\langle\delta\phi_{\alpha}(\eta_{1},\textbf{x}_{1})\delta\phi_{\beta}(\eta,\textbf{x}_{2})\rangle = \rho_{\Lambda}^{-1}\ \delta^{\nu}\ _{\mu}c^{\mu}\ _{\nu}
\end{equation}
\eqref{eq:final} is the probability amplitude of the two-point correlation function $C$ considered in \eqref{eq:cor} for which the WI was derived, which on analysis and evaluation has proved that correlation functions is quantum field theory are a probe to understand the fundamental laws and pertaining to inflationary paradigm, the quantum fluctuations follow a pattern as dictated by the symmetry constraints, reflected in the algebra followed by the WI in \eqref{eq:resf}, hence evolve on \textit{De Sitter} space-time as the background.
\chapter{Discussion of Results}
The Ward Identity ( WI ) for the two-point function in \eqref{eq:wi} contains the term $L_{\xi}\mathcal{L}$ which is the N\"{o}ether current for our theory, that remains conserved on  dragging the Lagrangian density along the Killing vectors, as a consequence of which \eqref{eq:cons} was obtained, whose validity depends on $\partial_{\mu}T^{\mu}\ _{\nu} = 0$. This clearly establishes the robustness of WI for doing model-independent analysis. The conservation of energy-momentum tensor is the physical reason for WI to obey \textit{closure property}. From \eqref{eq:wi}, there are set of 10 identities corresponding to the generator of isometry shown in \ref{tab:gen}, which can equally be derived by taking the commutation relations following \ref{tab:lie} acting on \eqref{eq:cor}, hence these set of identities are like elements forming a Lie group, which follows Lie algebra as proved in \eqref{eq:res}-\eqref{eq:resf} whose structure constants are the permutation tensors, hence \textit{isomorphic} to $so(3)$. Based on the explanation given in Section \ref{sec:sym}, Eqn\eqref{eq:so} and from Eqns.\eqref{eq:res}-\eqref{eq:resf}, the two-point correlation function for the quantum fluctuations belong to the de Sitter space-time. Their evolution is given by the equation of motion obtained in \eqref{eq:eom}, which had two cases \eqref{eq:time}-\eqref{eq:space}. Eqn.\eqref{eq:space} relates mass and potential for the Lagrangian density considered in \eqref{eq:inf}. Following \eqref{eq:cons}, up to the level of contact terms it is appropriate to relate the energy-momentum tensor to the Einstein tensor, moreover as proved that the two-point correlation function evolves on de Sitter space-time as the background which is maximally symmetric, Ricci tensor is proportional to the metric \eqref{eq:gmn}  and energy-momentum tensor corresponds to the cosmological constant. This geometrical approach helps in evaluating the probability amplitude of two-point function shown in \eqref{eq:final}. It is to be noted that research work\cite{conf,ope} using this approach is in vogue, but the work I have done differs from it, on the basis of the fact that Lagrangian density for the multi-field inflation in my case is very generic and consists of cubic terms in the interaction. I have derived the Ward Identity from the first principles using path integral method involving quantum fluctuations as fields, in order to know the background. The generators correspond to Killing vectors and not conformal ones. I have done a bottom-up approach in my analysis.
\chapter{Conclusion}
From my work I conclude that, given a correlation function one can find out the corresponding nature of space-time by deriving the Ward Identity from first principles and know its Lie algebra by finding out the structure constants. Ward Identity was derived from path integral formalism for the two-point correlation function corresponding to Lagrangian density for multi-field inflation. WI was subjected to analysis and proved that it follows a Lie algebra, with permutation tensor $(\epsilon_{abc})$ as structure constants, hence isomorphic to $so(3)$. Therefore, quantum fluctuations during the inflationary epoch evolved on de Sitter space-time as the background. It is already known that inflationary paradigm supports de Sitter space-time for its success, but this derivation and analysis on WI establishes it using field theoretic methods and is further extensible for any model-independent analysis. The closure of WI depends on equation for conservation, $\langle\partial_{\mu}T^{\mu}\ _{\nu}\delta\phi_{\alpha}\delta\phi_{\beta}\rangle = 0$, exploiting which an equation of motion was derived and as a special case a relation between mass and potential was obtained. De Sitter space-time being maximally symmetric, a geometrical approach was used up to the level of contact terms and the probability amplitude of the two-point correlation function $C$ was calculated. Hence, correlation functions in quantum field theory is an efficient probe for understanding the fundamental laws.
\newpage
\appendix
\chapter{Christoffel Symbols}\label{app:sym}
Christoffel symbols are given as
\begin{equation}
\Gamma^{\sigma}_{\mu\nu} = \frac{1}{2}\ g^{\sigma\gamma}\ (g_{\gamma\mu,\nu} + g_{\gamma\nu,\mu} - g_{\mu\nu,\gamma})\ .
\end{equation}
As the metric is symmetric with no off-diagonal terms, one can make appropriate choices for $\mu$ and $\nu$, such that when $\mu = \nu = \eta$, non-vanishing terms involve $\gamma = \eta$ and $\sigma = \eta$
\begin{equation}
\Gamma^{\eta}_{\eta\eta} = \frac{1}{2}\ g^{\eta\eta}(g_{\eta\eta,\eta}) = \left(\frac{a'}{a}\right)
\end{equation}
where $g_{\eta\eta} = a^{2}(\eta)$ and $g^{\eta\eta} = \frac{1}{a^{2}}$. Following this, setting $\mu = i$, $\nu = j$ and $\mu = \eta$, $\nu = j$ we get
\begin{equation}
\Gamma^{i}_{\eta j} = \Gamma^{i}_{j\eta} = \left(\frac{a^{'}}{a}\right)\ \delta^{i}_{j},\ \Gamma^{\eta}_{ji} = \Gamma^{\eta}_{ij} = \left(\frac{a^{'}}{a}\right)\ \delta_{ij}
\end{equation}
The rest of it are vanishing 
\begin{equation}
\Gamma^{i}_{ij} = \Gamma^{i}_{\eta\eta} = \Gamma^{\eta}_{\eta i} = 0\ .
\end{equation}
\chapter{Killing Vectors}\label{app:kill}
Consider Killing equation
\begin{equation}
\nabla_{\mu}\xi_{\nu} + \nabla_{\nu}\xi_{\mu} = 0\ .
\end{equation}
Expanding the covariant derivative
\begin{equation}
g_{\nu\sigma}(\partial_{\mu}\xi^{\sigma} + \Gamma^{\sigma}_{\mu\gamma}\xi^{\gamma}) + g_{\mu\sigma}(\partial_{\nu}\xi^{\sigma} + \Gamma^{\sigma}_{\nu\gamma}\xi^{\gamma}) = 0\ .
\end{equation}
From the results obtained in Appendix \ref{app:sym}, one can choose temporal and spatial coordinates accordingly for $\mu$ and $\nu$, such that when $\mu = \nu = \eta$
\begin{equation}
\nabla_{\eta}\xi^{\eta} = \partial_{\eta}\xi^{\eta} + \Gamma^{\eta}_{\eta\eta}\xi^{\eta} = 0
\end{equation}
which on solving gives
\begin{equation}
\xi^{\mu} \equiv \left(\frac{1}{a},0,0,0\right)
\end{equation}
hence the time translational isomerty is given as 
\begin{equation}
T = \frac{1}{a}\partial_{\eta}\ .
\end{equation}
Setting $\mu = i$, $\nu = i$ corresponds to 3-Spatial isometry
\begin{equation}
S_{i} = \partial_{i}
\end{equation}
followed by choosing $\mu = \eta$, $\nu = i$ and $\mu = i$, $\nu = j$ which gives Lorentz boost and angular momentum operator as Killing vectors  respectively
\begin{equation}
L_{i} = x^{i}\partial_{\eta} + \eta \partial_{i},\ J_{k} = -i \left(x^{i}\partial_{j}-x^{j}\partial_{i}\right)\ .
\end{equation}
\chapter{Riemann Tensor}
Riemann tensor is given by
\begin{equation}
R^{\alpha}\ _{\beta\mu\nu} = \Gamma^{\alpha}_{\nu\beta,\mu}-\Gamma^{\alpha}_{\mu\beta,\nu}+\Gamma^{\delta}_{\nu\beta}\Gamma^{\alpha}_{\mu\delta}-\Gamma^{\delta}_{\mu\beta}\Gamma^{\alpha}_{\nu\delta}
\end{equation}
Making appropriate substitutions for the indices,
\begin{equation}
\begin{split}
R^{\eta}\ _{x\eta x} & = \Gamma^{\eta}_{xx,\eta}-\Gamma^{\eta}_{\eta x,x}+\Gamma^{\delta}_{xx}\Gamma^{\eta}_{\eta\delta}-\Gamma^{\delta}_{\eta x}\Gamma^{\eta}_{x \delta} \\
                     & = \partial_{\eta}\left(\frac{a'}{a}\right) \\
                     & = R^{\eta}\ _{y \eta y} = R^{\eta}\ _{z \eta z}\ .
\end{split}
\end{equation}
Also,
\begin{equation}
R^{x}\ _{\eta x \eta} = g_{\eta\eta}R^{x \eta}\ _{x \eta} = g_{\eta\eta}R^{\eta x}\ _{\eta x} = g_{\eta\eta}g^{xx}R^{\eta}\ _{x \eta x} = - \partial_{\eta}\left(\frac{a'}{a}\right)
\end{equation}
and the Ricci tensor is
\begin{equation}
R_{\eta\eta} = R^{x}\ _{\eta x \eta} + R^{y}\ _{\eta y \eta} + R^{z}\ _{\eta z \eta} = -3\partial_{\eta}\left(\frac{a'}{a}\right)
\end{equation}
Similarly 
\begin{equation}
R^{x}\ _{yxy} = \Gamma^{\eta}\ _{yy}\Gamma^{x}\ _{x\eta} = \left(\frac{a'}{a}\right)^{2} = R^{x}\ _{zxz}
\end{equation}
and 
\begin{equation}
R^{y}\ _{xyx} = g_{xx}R^{yx}\ _{yx} = g_{xx}R^{xy}\ _{xy} = g_{xx}g^{yy}R^{x}\ _{yxy} = \left(\frac{a'}{a}\right)^{2}
\end{equation}
\begin{equation}
R^{z}\ _{xzx} = R^{x}\ _{zxz} = \left(\frac{a'}{a}\right)^{2}\ .
\end{equation}
The Ricci tensor is 
\begin{equation}
R_{xx} = R^{\eta}\ _{x \eta x} + R^{y}\ _{xyx} + R^{z}\ _{xzx} = \partial_{\eta}\left(\frac{a'}{a}\right) + 2\left(\frac{a'}{a}\right)^{2} = R_{yy} = R_{zz}
\end{equation}
Ricci scalar is given as
\begin{equation}
R \equiv g^{\mu\nu}R_{\mu\nu} = -\frac{6}{a^{2}} \left[\partial_{\eta}\left(\frac{a'}{a}\right)+\left(\frac{a'}{a}\right)^{2}\right]
\end{equation}
\chapter{Definitions}
1. Killing vector: Consider\cite{ano} a vector field $\xi = \xi^{\mu}\partial_{\mu}$ on a space-time manifold $(M,g)$. Let the infinitesimal displacement $\epsilon \xi$ be the generator of an isometry - a transformation
\begin{equation}
x^{\mu} \rightarrow x^{\mu} + \epsilon \xi^{\mu}
\end{equation}
such that
\begin{equation}
g_{\alpha \beta}\frac{\partial(x^{\alpha} + \epsilon \xi^{\alpha})}{\partial x^{\mu}}\frac{\partial(x^{\beta} + \epsilon \xi^{\beta})}{\partial x^{\nu}} = g_{\mu\nu}(x)
\end{equation}
which preserves the metric. Expanding to first order in $\epsilon$
\begin{equation}\label{eq:xi}
\xi^{\lambda}\partial_{\lambda}g_{\mu\nu}(x) + g_{\alpha\nu}(x)\partial_{\mu}\xi^{\alpha} + g_{\mu\beta}(x)\partial_{\nu}\xi^{\beta} = 0
\end{equation}
gives the Killing equation, whose solution $\xi$ is the Killing vector. \eqref{eq:xi} can be rewritten in terms of Lie derivative 
\begin{equation}
L_{\xi}g_{\mu\nu}(x) = \nabla_{\mu}\xi_{\nu} + \nabla_{\nu}\xi_{\mu} = 0 \ .
\end{equation}
2. Flow: For\cite{ano} a vector field $\xi$ on a manifold $M$, a curve $\sigma(t,x)$ whose tangent vector is $\xi$ passing through $x$ at a time $t$ is called as an integral curve. The map defined by the integral curve $\sigma : \textbf{R} \times M \rightarrow M$, is called as the flow of the vector field $\xi$. \\
\\
3. Lie derivative of a scalar: For\cite{ano} a scalar $\phi$, its differentiation with respect a vector field $\xi$ is defined as its Lie derivative 
\begin{equation}
L_{\xi}\ \phi = \lim_{t \rightarrow 0}\frac{1}{t}\left[\phi(\sigma_{t}(x))-\phi(x)\right] = \lim_{t \rightarrow 0}\frac{1}{t}\left[\phi(x^{\mu}+t\xi^{\mu}(x))-\phi(x)\right]
\end{equation}
hence
\begin{equation}
L_{\xi}\ \phi = \xi^{\mu} \frac{\partial}{\partial x^{\mu}}\ \phi = \xi^{\mu}\partial_{\mu}\phi
\end{equation}
which geometrically describes the moving or dragging of scalar $\phi$ along the flow $\sigma$ generated by the vector field $\xi$.\\
\\
4. Lie group: A group $G$ spanned\cite{pes} by elements $g$ arbitrarily close to the identity, such that the general element can be obtained by repeated action of the infinitesimal elements, such that
\begin{equation}
 g(\alpha) = 1 + i \alpha^{a}\ X^{a} + \mathcal{O}(\alpha^{2})
\end{equation}
where $\alpha^{a}$ is the group parameter and $X^{a}$ the generators of the group. Such a continuous group is called a Lie group.\\
\\
5. Lie algebra: A vector space spanned\cite{pes} by the set of generators $X^{a}$ of the Lie group $G$, obeying the commutation relation
\begin{equation}
[X^{a}, X^{b}] = i\ f^{ab}_{c}\ X^{c}
\end{equation}
where contants $f^{ab}_{c}$ are called structure constants, is called as Lie algebra.\\
\\
6. Embedding: Given\cite{ano} a Lie group and its associated Lie algebra, there are structures contained in them, called as sub-groups and sub-algebras obeying the same rules as defined for the former. These sub-structures are said to be embedded (contained); a smooth map which is an injection (one-to-one) and an immersion.

\newpage
\chapter*{Acknowledgements}
I acknowledge my supervisor Dr. David Seery, Prof. Andrew R. Liddle and faculty members of Department of Physics and Astronomy, University of Sussex. I take this opportunity to pay my respect to Prof. John Clive Ward (August 1, 1924 - May 6, 2000) and Prof. Yasushi Takahashi (December 12, 1923- ).

\end{document}